\documentclass[english,10pt, a4paper]{article}
\usepackage[english]{babel}
\usepackage[utf8]{inputenc}
\usepackage{amsmath}
\usepackage{amsfonts}
\usepackage{wasysym}
\usepackage{graphicx}
\usepackage[margin=20mm,top=20mm,bottom=20mm]{geometry}
\usepackage{float}
\usepackage{algorithm, comment}

\newcommand{\real}{\mathbb{R}}
\newcommand{\mm}[0]{\,{\textrm{mm}}}
\newcommand{\s}[0]{\,{\textrm{s}}}
\newcommand{\meas}[1]{\hat{#1}}
\newcommand{\accurate}[1]{\bar{#1}}
\newcommand{\fast}[1]{{#1}}

\begin{document}

\title{Model reduction in acoustic inversion by artificial neural network}

\author{Janne Koponen${}^{a,b}$, Timo L\"ahivaara${}^{a}$, Jari Kaipio${}^{c}$, Marko Vauhkonen${}^{a}$\\
\normalsize ${}^{a}$Department of Applied Physics, University of Eastern Finland, Kuopio, Finland\\
\normalsize ${}^{b}$School of Engineering and Technology, Savonia University of Applied Sciences, Kuopio, Finland\\
\normalsize ${}^{c}$Department of Mathematics, University of Auckland, Auckland, New Zealand}

\maketitle

\section*{Abstract}

In ultrasound tomography, the speed of sound inside an object is estimated based on acoustic measurements carried out by sensors surrounding the object.  An accurate forward model is a prominent factor for high-quality image reconstruction, but it can make computations far too time-consuming in many applications. Using approximate forward models, it is possible to speed up the computations, but the quality of the reconstruction may have to be compromised. In this paper, a neural network -based approach is proposed, that can compensate for modeling errors caused by the approximate forward models. The approach is tested with various different imaging scenarios in a simulated two-dimensional domain. The results show that with fairly small training datasets, the proposed approach can be utilized to approximate the modelling errors, and to significantly improve the image reconstruction quality in ultrasound tomography, compared to commonly used inversion algorithms.

\section{Introduction}

In ultrasound tomography, speed of sound (SOS) distribution in an object is estimated based on acoustic measurements. Mathematically, the reconstruction of SOS is a nonlinear inverse problem. Many different techniques have been developed to solve this problem, for example, algebraic reconstruction methods \cite{gordon1970algebraic, kak1979computerized, andersen1984simultaneous}, adjoint methods \cite{tarantola1988theoretical, fichtner2006adjoint1, fichtner2006adjoint2, plessix2006review}, and artificial neural networks (ANN) \cite{zakharia1996estimation, thompson2003inversion, daliakopoulos2005groundwater, lahivaara2018deep}. Algebraic reconstruction methods are typically computationally quite light methods to solve the inverse problem. On the other hand, a fairly large number of sensors are required to achieve high-quality reconstructions. Adjoint methods, on the other hand, can produce high-quality reconstructions with a smaller number of measurements. However, the adjoint methods are computationally more demanding, and thus they can be impractical in some cases. The ANN-based inversion methods are typically computationally less intensive, but require a large number of training samples, and therefore these methods can be infeasible in some applications.  

Solving the acoustic inverse problem with an adjoint method, a large number of corresponding forward problems need to be solved. Computational requirements of computing the reconstruction can be reduced by using computationally less expensive forward models. However, these models are typically less accurate compared to complete models, and hence, the quality of the reconstruction tends to be lower. Various methods have been developed to reduce the accuracy requirements of the forward model while preserving the accuracy of the inverse solution. These methods include, for example, the Bayesian Approximation Error method (BAE) \cite{kaipio2006statistical, damien2013bayesian32}, correction methods \cite{10.1093/gji/ggu057, lunz2020learned}, and Hybrid Algorithm for Robust Breast Ultrasound Tomography (HARBUT) \cite{huthwaite2011high}. For example in BAE, the modeling errors are approximated in the statistical framework by Gaussian distributed random variables, and the reconstruction is achieved by computing a {\it maximum a posteriori} estimate of the parameters. 

Artificial neural networks are a computational model that is inspired by neurobiological systems. The ANN has been used in numerous applications, such as image recognition \cite{zoph2018learning, rawat2017deep}, natural language processing \cite{goldberg2016primer}, economy \cite{moody1995economic,angelini2008neural,li2010applications}, and medicine \cite{bate1998bayesian, ngo2019semisupervised}. Also, as mentioned earlier, the ANN has been utilized to solve inverse problems \cite{bianco2019machine}. In the SOS inverse problem, the standard approach is to train the ANN with a large set of simulated SOS distributions and the corresponding synthetic acoustic measurements. In the image reconstruction, the input to the network is the acoustic measurement data and the outcome is the desired SOS distribution.    

A drawback of common ANN approaches for solving inverse problems is the computationally expensive training phase, requiring tens of thousands of training datasets. This may require a notably large amount of computational resources already in the training phase. 
In this paper, we propose a novel image reconstruction approach that utilizes the ANN to create a numerical function that converts the measured signal to a corresponding low dimensional, surrogate model, signal.  This means, that the chosen ANN can be trained with a (fairly small) set of SOS distributions with the corresponding accurate acoustic data and the desired light (surrogate) model data. In the inverse solution phase, the measured data is converted to the surrogate model data with the trained ANN and the final image reconstruction is carried out using a standard adjoint approach, with the chosen surrogate model. In the rest of the paper, the proposed approach is called the Neural Network based Approximation Error Estimation method (NNAEE). 

The rest of the paper is organized as follows. The acoustic forward model is presented in Section \ref{sec:fw}, modelling errors and the ANN approach of using a surrogate model for solving the inverse problem is discussed in Sections \ref{sec:moderr} and \ref{sec:NNAEE}. In Section \ref{sec:numex}, the proposed method is demonstrated by numerical examples, and finally, in Section \ref{sec:concl} the conclusions are given. 

\section{The forward model}\label{sec:fw}

\begin{figure}[!ht]
  \center{
  \includegraphics[scale=1]{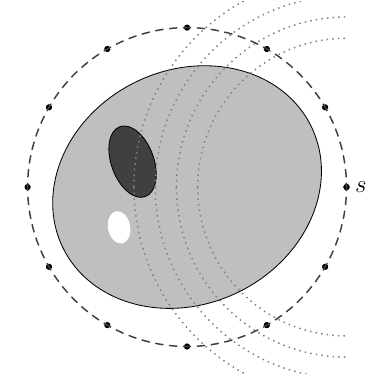}  
  }
  \caption{Schematic figure of a measurement setup in ultrasound tomography. The transmitter $s$ sends waveform, which propagates through the object. The propagated signal is measured by sensors located on the dashed circle.}
  \label{fig:schematic}
\end{figure}

In ultrasound tomography, $M_s$ transmitters and $M_r$ receivers are located around the region of interest $\Omega$. A schematic picture of the measurement setup is presented in Figure \ref{fig:schematic}. One transmitter at a time sends an acoustic signal, which propagates through the object. The propagated signal is measured by all the receivers simultaneously. 
Wave propagation is modeled by the wave equation in an unbounded domain, $x \in \real^2$ and $t> 0$, as

\begin{equation}
\label{eq:waveeq}
-\nabla^2 u + \frac{1}{c^2} \frac{\partial^2 u}{\partial t^2} = f \delta(x-x_{t}),
\end{equation}
with the initial conditions
\begin{equation*}
\left\{ \begin{array}{lcll}
u(x,0) & = & 0  \\
\frac{\partial}{\partial t} u(x,0) & = & 0,
\end{array} \right.
\end{equation*}
where $u=u(x, t)$ is pressure, $c=c(x)$ is the speed of sound (SOS), $f=f(t)$ and $f\delta(x-x_t)$ is a source term, where $\delta(x-x_t)$ is Dirac delta function, $t$ is time, and $x_{t}$ is a position of the transmitter. The measured data  is collected at fixed time instants over a certain measurement period. When collecting data at  $M_t$ time instants, a total of $ M_r \times M_s \times M_t$ data points are obtained in the measurement vector $\meas{y}$. With a fixed source term $f\delta(x-x_t)$,  the measured data depends only on the parameter $v(x):=\frac{1}{c^2(x)}$.

In practice, when numerical models are utilized, the unbounded computational domain must be truncated. The arising boundaries generate reflections into the modeled signal. There are several ways to reduce the reflections by implementing Absorbing Boundary Conditions (ABC). These include methods such as perfectly matching layers \cite{liu1997perfectly, berenger1994perfectly}, high-order boundary conditions \cite{hagstrom2004new, hagstrom2009complete} and absorbing layers near boundaries \cite{israeli1981approximation, semblat2011simple}. In addition, there are numerous methods to solve the wave equation, see for example \cite{lahivaara2008computational, rose1982time, yang2004optimal, manry1996fdtd, lines1999recipe, komatitsch2010accelerating, bojarski1982k, bojarski1985k}.  In this paper, a finite-difference time-domain (FDTD) solver is used to solve the forward problem. In addition, absorbing layers near the boundaries are utilized to implement the ABC. The absorbing layers are implemented by adding a damping term, $\eta \frac{\partial u}{\partial t}$, into left hand side of equation \eqref{eq:waveeq}. Inside of area surrounded by the sensors, $\eta = 0$, and $\eta > 0$ elsewhere.

\section{Model selections, inverse problem and modelling error}
\label{sec:moderr}

{In practice, the true physical measurement can be approximated by some high accuracy observation model}
\begin{equation}
\label{eq:accurate_y=Ax}
\accurate{y}=\accurate{A}(\accurate{v})+\accurate{e},
\end{equation}
where the model $\accurate{A}$ and the corresponding parameter vector $\accurate{v}$ approximate the true physical model as well as possible. This model, however, is computationally demanding and the use of it in inverse problem solution is not desirable. To overcome this, a less accurate forward model, also called as {\em surrogate model}, is chosen and is written in the form  
\begin{equation}
  \label{eq:inverseprob}
  \fast{y} = \fast{A}(\fast{v})+e,
\end{equation}
where $\fast{A}$ is {computationally light, and, in many cases less accurate,} forward model and $\fast{y}$ is the corresponding data vector. Here the parameter vector {$\fast{v} = P\accurate{v}$}, where the operator $P$ projects {$\accurate{v}$} to the lower dimensional basis of the surrogate model. 

\subsection{The proposed NNAEE approach}

In solving the inverse problems, the accuracy of the forward model and
its numerical implementation play a crucial role. In general, accurate
numerical models are expensive to compute, and on the other hand,
computationally lighter models are less accurate.  In the proposed
NNAEE approach we proceed as follows. Equation
{\eqref{eq:accurate_y=Ax}} can be written as

\begin{equation*}
\accurate{y}=\fast{A}(\fast{v}) +(\accurate{A}(\accurate{v})+\accurate{e}-\fast{A}(\fast{v}))=\fast{A}(\fast{v}) -\nu(\accurate{y}). 
\end{equation*}

This can further be written as
\begin{equation*}
    \fast{A}(\fast{v})=\accurate{y}+\nu(\accurate{y})=:\phi(\accurate{y}),
\end{equation*}
where {$\phi(\accurate{y})$ is the unknown mapping that converts the
  accurate model signal $\accurate{y}$ to a corresponding surrogate
  model signal $\fast{A}(\fast{v})=\fast{y}$. Because $\accurate{y}$
  is high quality approximation of real measurement signal, we can
  write $\fast{A}(\fast{v}) \approx \phi(\meas{y})$, where $\meas{y}$
  is real world measurement signal, and further
\begin{equation*}
    \fast{A}(\fast{v}) \approx \meas{y}+\nu(\meas{y})  =\phi(\meas{y}).
\end{equation*}
Mapping $\phi$ is extremely complex, and finding a closed form
approximation of the mapping would be difficult. On the other hand,
the universal approximation theorem states \cite{NIPS2017_7203,
  math7100992} that a large space of real-valued functions can be
approximated by ANN. Therefore, in the NNAEE approach, the mapping
$\phi$ is trained and approximated by an ANN. To compute the
reconstruction $v$ from the measured signal $\meas{y}$, the surrogate
numerical model (\ref{eq:inverseprob}) can be utilized without
compromising the quality, if the mapping $\phi$ is accurate enough.
In the current paper, we utilize an adjoint method, similar to
\cite{roy2010sound}, to solve the inverse problem. However, other
methods could also be used.

\section{The structure of the proposed approach}\label{sec:NNAEE}

In this section, the structure of the neural network including the autoencoder/-decoder phase is described in detail. Also, the algorithm of the proposed NNAEE approach is introduced.  

\subsection{Notations and parameters of the models used}

In the numerical tests later on, we utilize the following three different models
\begin{enumerate}
  \item {High quality model used to generate simulated measurement signals,} $\meas{A}(\meas{v})$
  \item High quality numerical model $\accurate{A}(\accurate{v})$
  \item Computationally light surrogate model $\fast{A}(\fast{v})$
\end{enumerate}
 
The first, simulated physical model mimics the real physical world. In simulations, the model utilizes finite-difference discretization in the spatial domain and leapfrog integration in the time domain. In addition, the measurement noise is simulated by Gaussian noise.

The second, accurate model is a high-quality numerical model of the physical system. The model contains, also, measurement error. However, in real world, the true measurement noise is rarely known, and hence we use a slightly different distribution for the measurement noise compared to the noise of the simulated physical model.

The third, the surrogate model, is a numerical model which is computationally light and is used in the final image reconstruction. The model utilizes coarse spatial discretization, and low-quality ABC.

\subsection{Autoencoder and autodecoder}

The measurement signal contains measurement noise, and in addition, the true information content of the measurement vector is less than the actual size of the signal. An autoencoder (AE) and autodecoder (AD) can be utilized to reduce the measurement vector dimension \cite{kramer1991nonlinear, hinton1994autoencoders}, and, in the same time, to suppress measurement noise \cite{vincent2008extracting, vincent2011connection}.

In the proposed NNAEE approach for each transmitter-receiver pair, we utilize a three-layer ANN for AE and a three-layer ANN for AD, presented in Figure \ref{fig:ae+ad}. This means that each measurement signal of length $M_t$ of one transmitter-receiver pair is reduced to size $M_p$ with the AE unit. The total length of the reduced dataset is thus $\real^{M_r \times M_s \times M_p}$. The AD unit is used to decode the  signal of one transmitter-receiver pair back to the original length $M_t$. In the simulation test cases, we have used $M_t=2363$ and the width of the information bottleneck is chosen as $M_p=50$. 

\begin{figure}[!ht]
\center {
  \includegraphics[width=0.4\linewidth]{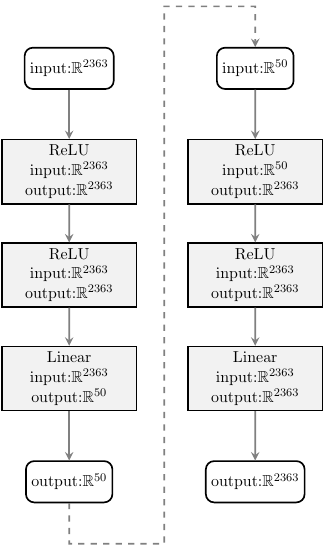}
}
\caption{Three layers implementing the autoencoder, mapping $\alpha$, on left, and three layers implementing the autodecoder, mapping $\beta$, on right. Layers 1 and 2 in both cases are dense layers using ReLU (rectified linear unit) activation, and the last layers are dense layers with linear activation. At the training phase, the autoencoder output is connected to the autodecoder input.}
\label{fig:ae+ad}
\end{figure}

We train the AE+AD by using both the surrogate model generated signals and accurate model generated signals. We utilize both of the signals because they are needed at a later stage for training the mapping $\phi$ and a large number of training {examples} decreases the risk of over-fitting. Because there are $M_s \times M_r$ transmitter-receiver pairs, there are in total $2 \times M_s \times M_r \times N_{\mathrm{train}}$ samples for AE+AD training. 
 Input of the AE+AD, denoted as $X=A_{s,r}^*(v^*)+e^*$, where $A_{s,r}^*(v^*)+e^*$ is either $\fast{A}(\fast{v})+\fast{e}$ or $\accurate{A}(\accurate{v})+\accurate{e}$, and
 $s,r$ is a transmitter-receiver pair, and $e^*$ is {the corresponding} measurement noise. The desired output of the AE+AD is $Y=A_{s,r}^*(v^*)$, that is, the same signal as the input signal but without the measurement noise. From now on, we denote AE as mapping $\alpha$, and AD as mapping $\beta$.

\begin{figure}[!ht]
\center {
  \includegraphics[width=\textwidth]{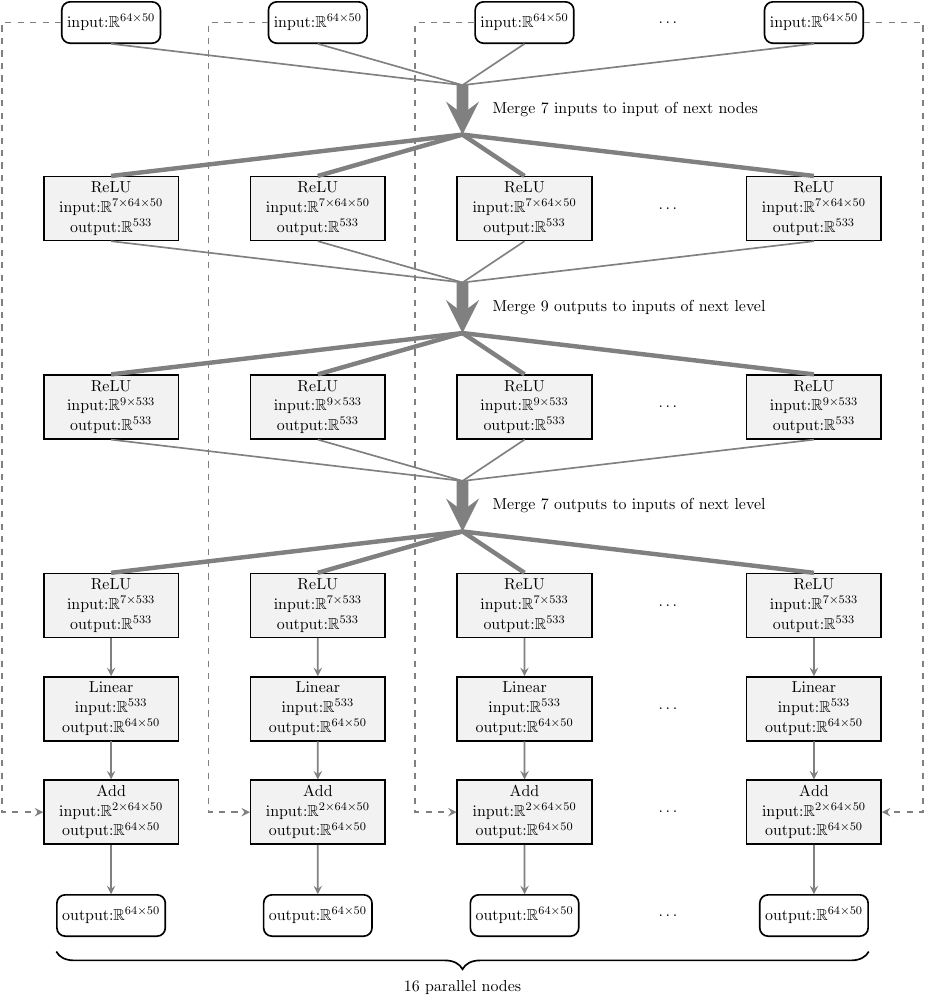}
}
\caption{A partial figure of the ANN approximating $\varphi$. Each input contains all the measurements of one transmitter, compressed with the AE. The input layer contains $M_s=16$ inputs, each size of $M_r \times M_p = 64 \times 50$. Each ReLU-net at the first layer inputs data from 7 inputs, each net at the second layer utilizes data from 9 outputs of the first layer etc. At the last layer, each input is added to the corresponding output of the second last layer.}
\label{fig:phi_ann}
\end{figure}

\subsection{Mapping $\phi$}

Let us define a composite mapping {$\phi(\accurate{y})=\beta(\varphi(\alpha(\accurate{y})))$,} where $\varphi$ is approximated by ANN, shown in Figure \ref{fig:phi_ann}, and $\alpha$ and $\beta$ are the AE and AD, respectively. To keep notation simple, we utilize here $M_s \times M_r$ parallel AE and AD, one for each transmitter-receiver pair. The input and output vectors of $\varphi$, both are in $\real^{M_s \times M_r \times M_p}$. Using multiple layers of dense networks of such size is infeasible.  Decreasing the number of neuron connections it is possible to create an ANN which does not require a large amount of GPU memory and can be implemented in a standard desktop computer with, for example, 8 GB of GPU memory. Decreasing the connections in the proposed NNAEE approach is implemented by forming layers that contain parallel dense networks, with the feature that all the networks are not connected to all neurons on the consecutive layers, see Figure \ref{fig:phi_ann}. Between each layer, selected outputs of earlier level are merged for the input of the next level networks. Each subnet input of the next level has unique set of earlier outputs. The network contains three layers of dense networks using rectified linear units (ReLU), and one layer with linear activation function. The last layer sums the input and output of the linear units forming the output of $\varphi$.

\subsection{The NNAEE image reconstruction algorithm}

Collecting steps defined in earlier subsections, Algorithm \ref{alg:nnaee} can be obtained. The algorithm has two phases; one is done beforehand, only once, and the other phase does the reconstruction and utilizes the actual measurement signal. As the first step is done only once, the step can utilize large amounts of computer resources. In practice, the step can be computationally quite demanding, because the training of an ANN can require a large amount of training data. The second step, on the other hand, utilizes the trained ANN, which does not require an excessive amount of computing power, and most of the computing power at this phase is utilized for solving the actual inverse problem with the adjoint method.

\begin{algorithm}[!ht]
\textbf{A. Once, before any measurements}
\begin{enumerate}
  \item Create training measurement signals
    \begin{enumerate}
      \item Generate SOS fields representing the assumed true distributions 
      \item Generate simulated measurement signals with the accurate model $\accurate{A}(\accurate{v})$
      \item Generate simulated measurement signals with the surrogate model $\fast{A}(\fast{v})$
    \end{enumerate}
  \item Train AE/AD 
  \item Encode training signals by AE -- Output is the training dataset
  \item Train ANN representing the mapping $\varphi$ using encoded training dataset to form $\phi(\cdot)=\beta(\varphi(\alpha(\cdot)))$.
\end{enumerate}
\ \\
\ \\
\textbf{B. With the actual (or simulated) measurement data}
\begin{enumerate}
  \item Perform a measurement (or simulate with $\meas{A}(\meas{v})$) 
  \item Convert signal to corresponding surrogate model signal with the mapping $\phi$
  \item Solve the inverse problem (reconstruct the unknown $v$ field) using the converted signal and the chosen surrogate model.
\end{enumerate}

\caption{NNAEE}
\label{alg:nnaee}
\end{algorithm}

\section{Numerical tests}\label{sec:numex}

In this section, we demonstrate the NNAEE method by investigating the dependency of the quality of reconstruction on the size of the training dataset. We also compare the proposed method to an accurate model reconstruction and a conventional reconstruction method (CRM). In the accurate model reconstruction, we use the model $\accurate{A}(\accurate{v})$ and in CRM the reconstruction is obtained using the approximate model $\fast{A}(\fast{v})$. In addition, the method is compared to a direct ANN-based inversion. The ANN-based inversion is implemented with a similar type of ANN as presented in Figure \ref{fig:phi_ann}. Comparison utilizes Root Mean Square Error,
\begin{equation*}
    {\rm RMSE} (v, w) = \sqrt{\frac{1}{N_{\mathrm{grid}}^2} \sum_{i=1}^{N_{\mathrm{grid}}}\sum_{j=1}^{N_{\mathrm{grid}}} (v_{i, j}-w_{i, j})^2}\ \ ,
\end{equation*}
where $v$ and $w$ are $N_{\mathrm{grid}} \times N_{\mathrm{grid}}$ SOS fields.

Training the ANN requires a large number of training data, and therefore using real measurements to train the ANN is not feasible. The training dataset and validation dataset are generated using a high quality numerical model $\accurate{A}(\accurate{v})$, and the surrogate numerical model $\fast{A}(\fast{v})$. The training input data $X=\alpha(\accurate{A}(\accurate{v}))$, and the preferred output $Y=\alpha(A(v))$, where $v=P\accurate{v}$. In the simulations, the number of transmitters and receivers were $M_s=16$ and $M_r=64$, respectively.  

We use the following sizes of the training datasets: $N_{\mathrm{\mathrm{train}}} \in \{200,\, 400,\, 800,\, 1200,\, 1800,\, 2500,\, 5000,\, 10000,$ $25000,\, 50000 \}$ to see how the size of the training set affects the accuracy of the reconstruction. Each training/validation dataset item is generated as shown in Algorithm \ref{alg:trainset}. The size of a validation dataset was 200 samples, and the size of a test dataset was $20$ samples. The validation dataset was used to tune the hyperparameters of the network and training. The test dataset was not used at the training phase. It was only used to test the performance of the proposed method.

\subsection{Model parameters}\label{sec:mod_para}

The parameters of the models are presented in Table
\ref{tbl:model_parameters}.  All models utilize FDTD leapfrog solver with $\Delta t=10^{-7}\s$ time step to solve the partial differential equation. The spatial grid resolution $\Delta x$ and the grid size $N_{\mathrm{grid}}$ change over models, as well as the quality of the ABC. {Transmitters and receivers are located on a circumreference with the radius of $100\mm$, and the radius of the region of interest is $80\mm$. }In AE+AD training phase an input signal contains Gaussian random noise with variance of $\sigma^2_e$, which is added to the measurement signals generated by the accurate and surrogate models. On the other hand, a desired output signal does not contain noise. In training of the ANN, noisy accurate simulated signal packed with AE is used along with the desired noiseless surrogate model signal. In reconstruction and test phases, the "true" simulated signal with added noise is utilized.

\begin{table}[!ht]
  \caption{Parameters of the models used in the study. The notations are explained in the text. Signals generated by accurate model and surrogate model are utilized, depending on the case, with or without simulated measurement noise. ABC layer thickness is the minimum thickness of the absorbing layer near the boundary, and $\lambda$ is the wavelength at the frequency of $50000\,\rm{Hz}$ with the SOS of $1500\,\rm{m/s}$.}
  \label{tbl:model_parameters}
  \center{
\begin{tabular}{|c|c|c|c|}
\hline
 & Simulated physical & Accurate & Surrogate\\
 & model & model & model\\
& $\meas{y}=\meas{A}(\meas{v})+\meas{e}$ & $\accurate{y}=\accurate{A}(\accurate{v})+\accurate{e}$ & $\fast{y}=\fast{A}(\fast{v})+\fast{e}$\\
\hline
\hline
$\Delta x$ & $0.2557\mm$& $0.2637\mm$ & $0.8398\mm$ \\
\hline
$N_{\mathrm{grid}}$ & $1056\times1056$ &  $1024\times1024$ &  $256\times256$\\
\hline
ABC layer   & $35\mm$ & $35\mm$ & $7.5\mm$\\
thickness & $1.17\lambda$ & $1.17\lambda$ & $0.25\lambda$\\
\hline
$\sigma^2_e$ & $0.0006^2$ & $0.0005^2\, /\, -$ & $0.0005^2\, /\, -$\\
\hline
\end{tabular}
}
\end{table}

\begin{algorithm}[!ht]
\begin{enumerate}
    \item Generate a high quality SOS field $\accurate{v}$ with the {discretization} of $1024 \times 1024$ {points}.
    \begin{itemize}
        \item Background $c(x)$ is Gaussian, $E[c(x)]=1500\,\rm{m/s}$, cutting the values of $c(x)$ in $1100\,\rm{m/s}\dots \, 1800\,\rm{m/s}$, standard deviation of background SOS,  $\sigma_{\mathrm{background}}=91\,\rm{m/s}$. 
        \item On the background, three circular shaped inclusions are inserted with constant SOS values $c_i,\ i=1,\ldots,3$, where $c_i$ is Gaussian with  $E[c_i]=1500\,\rm{m/s}$ and $\sigma_{c_i}=50\,\rm{m/s}$.
    \end{itemize}
    \item Compute a low resolution SOS field. $\fast{v}=P\accurate{v}$ with the {discretization} of $256 \times 256$ {points}.
    \item Compute the simulated signal with the accurate model, $\accurate{y}=\accurate{A}(\accurate{v})+\accurate{e}$ with the source term.
    \begin{equation*}
    f(t)=\exp\left(\xi(t-t_0)^2\right)\, \sin(2 \pi f_0 t),
    \end{equation*}
    where $\xi=-10^{10}\,$s${}^{-2}$, $f_0=50000\,$Hz, and $t_0=0.000025\,$s. 
    \item Compute the simulated signal with the surrogate model, $\fast{y}=\fast{A}(\fast{v})$, without any noise.
    \item $\accurate{y}$ is used as an input and $\fast{y}$ is used as the desired output when training the ANN approximating $\varphi$.
\end{enumerate}
\caption{Generating a training/validation dataset item}
\label{alg:trainset}
\end{algorithm}

\subsection{Test dataset}

Twenty test cases were generated to validate the method. Simulated measurement signals were computed using the simulated physical model $\meas{y}=\meas{A}(\meas{v})+\meas{e}$. The SOS fields used in the test cases are significantly different from the SOS fields used at the training set, see Figure \ref{fig:sosfields}. For example, SOS fields in the validation set contain star-shaped inclusions while in the training set there are only circular inclusions. The background variance of the sets are different, in test dataset $\sigma_{\mathrm{background}}=23\,\rm{m/s}$. In addition, the variance of the measurement error is different in the training dataset.

\begin{figure}[!ht]
\begin{flushleft}  
  \includegraphics[scale=0.5]{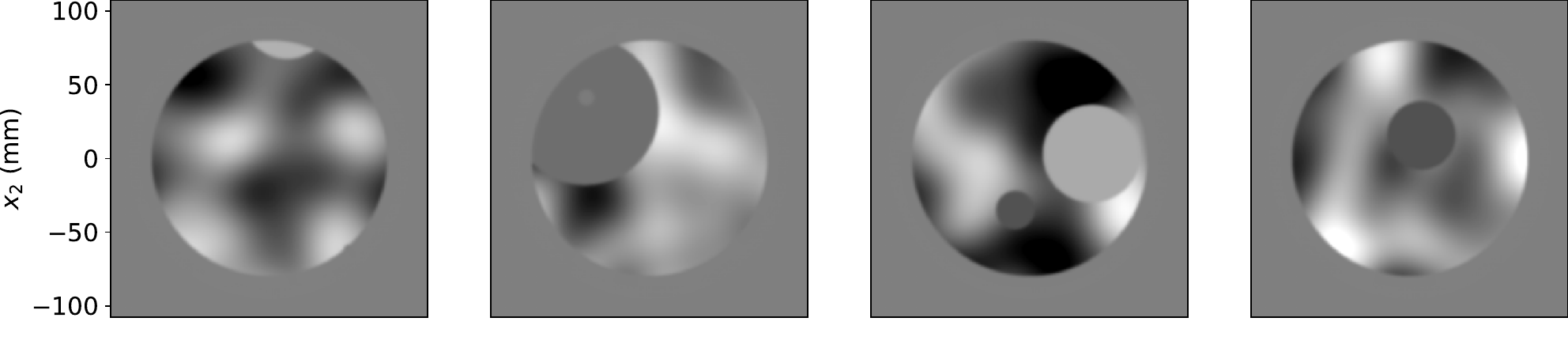}
  \includegraphics[scale=0.5]{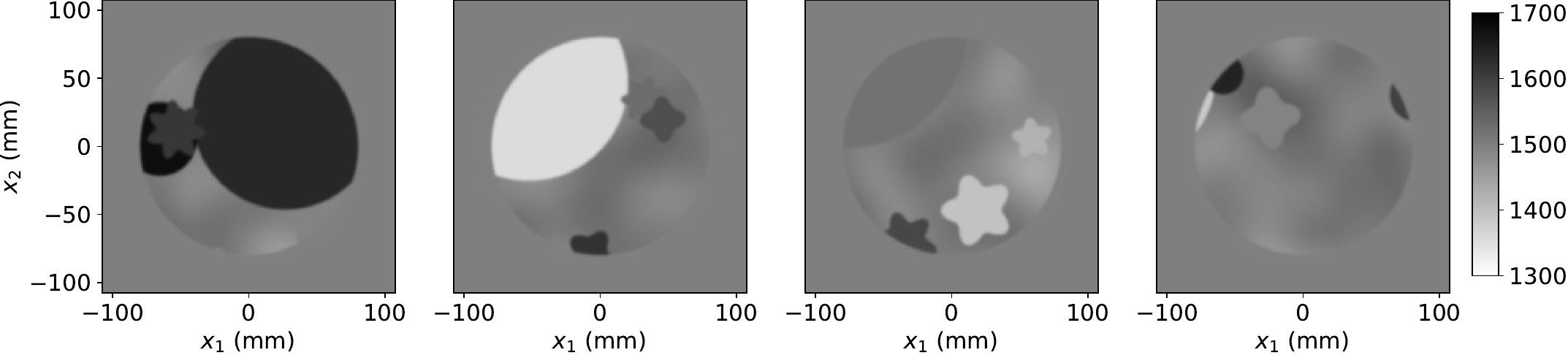}
  \end{flushleft}
  \caption{Examples of the used SOS fields. The upper row contains SOS fields drawn from the training set, and the lower row contains SOS fields drawn from the validation set. {The SOS fields are given in $\rm m/\s$.} }
  \label{fig:sosfields}
\end{figure}

\subsection{Results}

In the NNAEE method, we trained the ANN to obtain a mapping $\phi$ that converts the (simulated) measurement signals to corresponding signals of the surrogate model.  Figure \ref{fig:simusignals} shows two examples of the performance of the mapping.  The mean value of the error between the true signal and the surrogate model signal is 17.75, whereas when using the ANN converted signal, the average error goes down to 1.2. %

Examples of the reconstructions and the corresponding error images are presented in Figure \ref{fig:reco5000}
and the detailed results of the errors can be found in Figure \ref{fig:learning_curves} and Table \ref{tbl:result}. The method is seen to improve the quality of the reconstruction significantly, even with a relatively small size training dataset. For example, with the training set of size 200, the reconstruction RMS-error decreases from over $49.5$, corresponding to a conventional reconstruction method (CRM), down to below $9.8$, which is about two times the error of the reconstruction using the accurate model. On the other hand, growing the training set does not improve the results after above $5000$ training samples.  Using a training dataset of size $5000$ decreases the RMS-error to less than $6.9$, however, even with $50000$ training samples the error stays around $6.8$. 

Using the method does not increase the reconstruction time compared to CRM. The reconstruction time of the NNAEE and CRM is approximately $25\,\rm{min}$, which is significantly lower than the reconstruction time when using the accurate forward model, approximately $4\,\rm{h}\ 30\,\rm{min}$. On the other hand, the proposed method can generate  reconstructions
of similar quality as the standard ANN-based inversion with a significantly smaller training dataset. For example, the RMS-error of the NNAEE  reconstruction with $N_{\mathrm{train}}=200$ is $9.76$, while the standard ANN inversion requires almost $N_{\mathrm{train}}=10000$ to reach the same quality.   

\begin{figure}[!ht]
  \center{
  \includegraphics[width=0.5\textwidth]{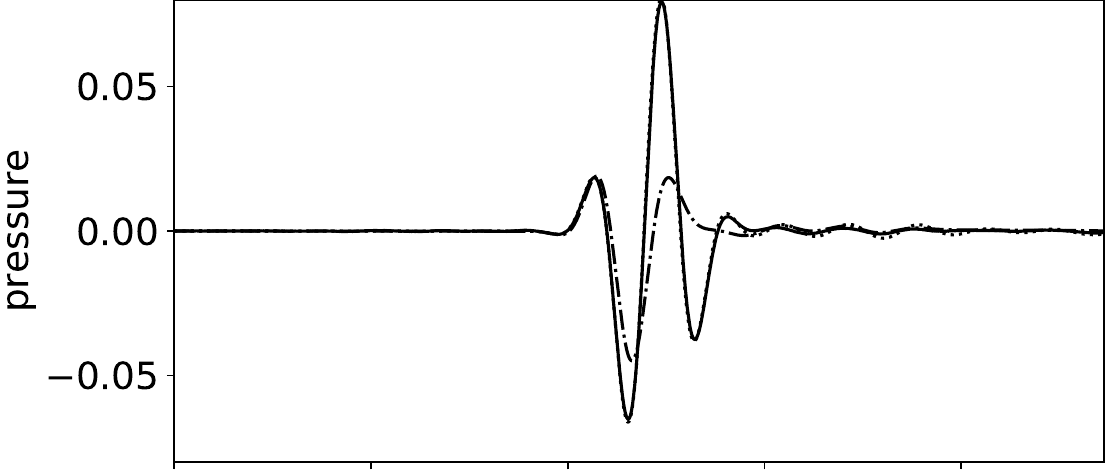}
  \includegraphics[width=0.5\textwidth]{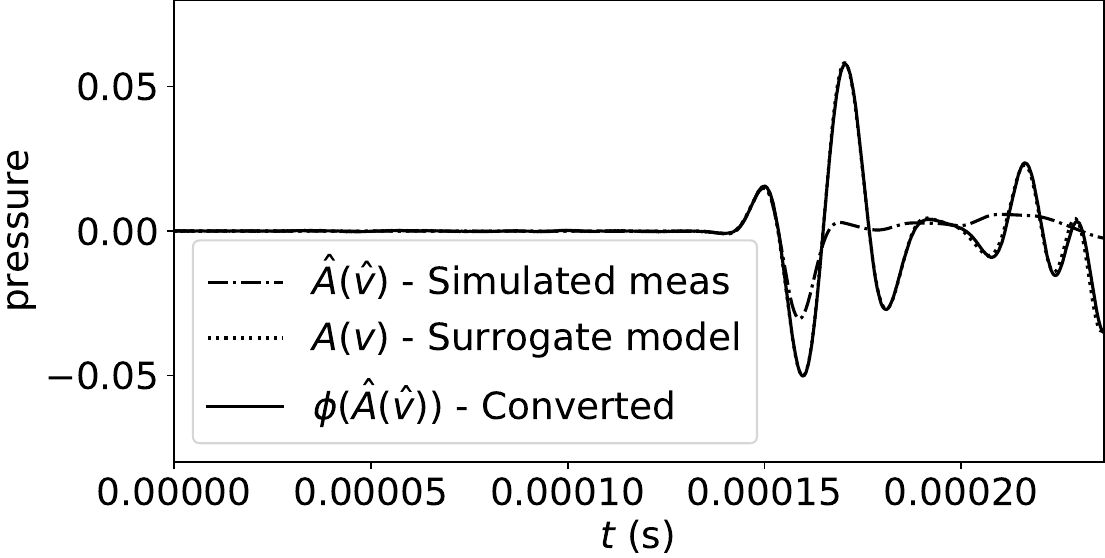}
  }
  \caption{Simulated measurement signals using different forward models. Dotted-dashed lines represent the simulated pressure signal without measurement noise, and dotted lines show the signals generated by the surrogate model with the same SOS field.  Solid lines, following almost exactly the surrogate model signal, show the ANN converted signal, which is used for computing the reconstruction. On the upper figure, the transmitter and receiver are at $90^\circ$ angle with respect to each other, and on the lower figure, the angle is $180^\circ$. {Note that the pressure amplitude is given in proportion to the amplitude of the source term $f(t)$.}} 
  \label{fig:simusignals}
\end{figure}

\begin{figure}[!ht]
  \center{
  \includegraphics[width=\textwidth]{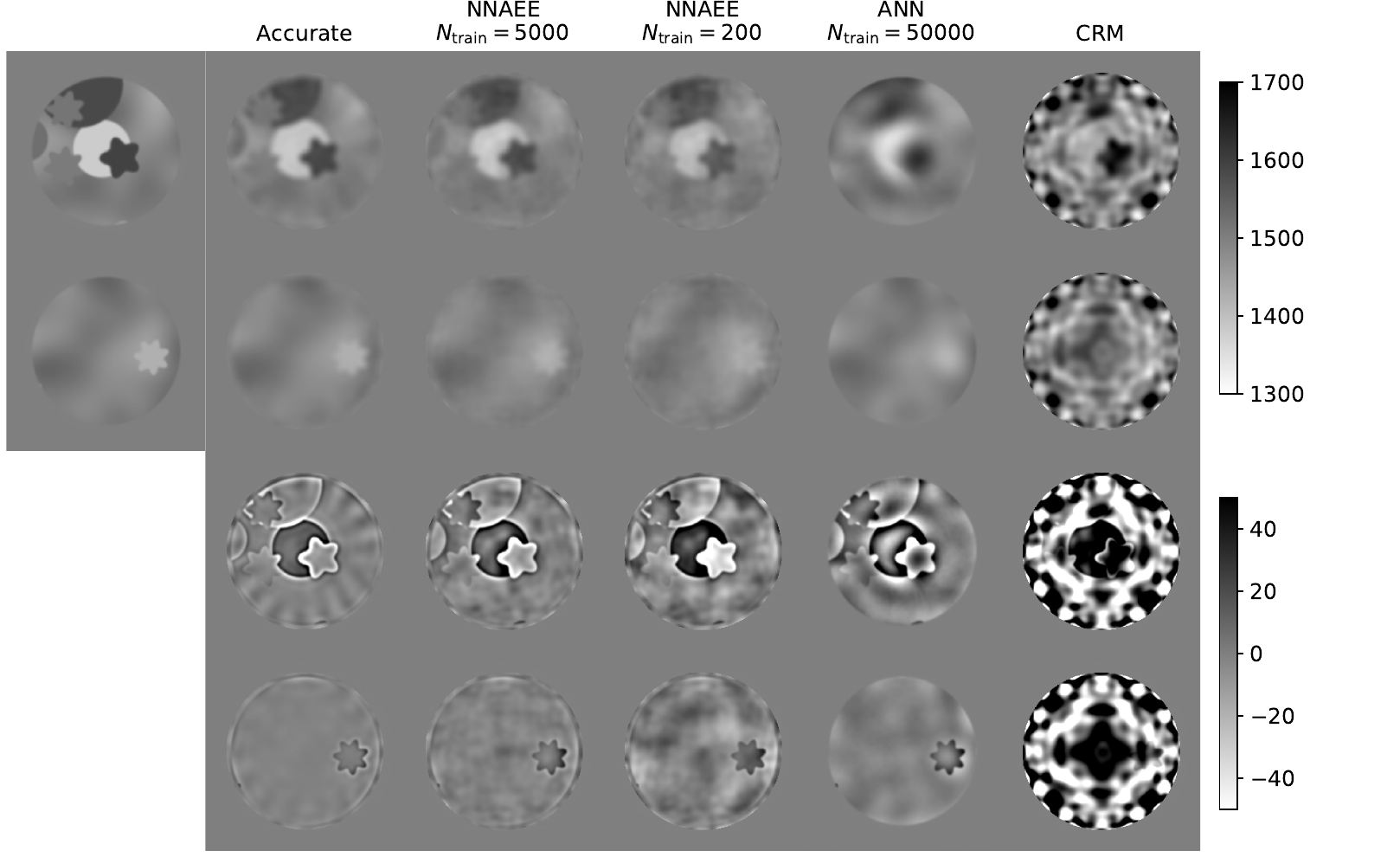}
  }
  \caption{Target is located on left, and reconstructions on two upper rows from left to right are computed with accurate model, NNAEE, with $N_{\mathrm{train}}=5000$, NNAEE, with $N_{\mathrm{train}}=200$, ANN based inversion, with $N_{\mathrm{train}}=50000$, and CRM. At lower rows there are the corresponding error fields. {The edge length of each rectangle is $215\rm{mm}$, and SOS and error fields  are in $\rm m/ \s$.}}
  \label{fig:reco5000}
\end{figure}          

\begin{table}[!ht]
 \caption{Average RMS-errors of the inversions using NNAEE and ANN inversion as a function of $N_{\mathrm{train}}$. The error decreases significantly first, and converges to the value of approximately 6.80 when $N_{\mathrm{train}}$ increases. Error of the reconstruction using the accurate model, $\accurate{A}(\accurate{v})$, is presented here as a lower limit reference value, which is possible to reach with a perfect $\phi$. In addition, error of the reconstruction using the surrogate model $A(v)$, is presented as the upper limit reference value.}
  \label{tbl:result}
  \center{
  \begin{tabular}{|c|c|c|}
    \hline
    $N_{\mathrm{train}}$ & NNAEE error ($\rm{m}/\rm{s}$)& ANN error ($\rm{m}/\rm{s}$)\\ \hline \hline
    200 & 9.76 & 21.63\\ \hline
    400 & 9.55 & 17.97\\ \hline
    800 & 8.61 & 15.77\\ \hline
    1200 & 8.13 & 15.25\\ \hline
    1800 & 7.32 & 13.66\\ \hline
    2500 & 6.99 & 13.63\\ \hline
    5000 & 6.89 & 12.09\\ \hline
    10000 & 6.89 & 9.70\\ \hline
    25000 & 6.80 & 7.92\\ \hline
    50000 & 6.80 & 8.02\\ \hline \hline
    Accurate & 5.03 & \\ \hline
    Surrogate & 49.54& \\ \hline
  \end{tabular}
  }
\end{table}

\begin{figure}[!ht]
  \center{
  \includegraphics[width=0.5\textwidth]{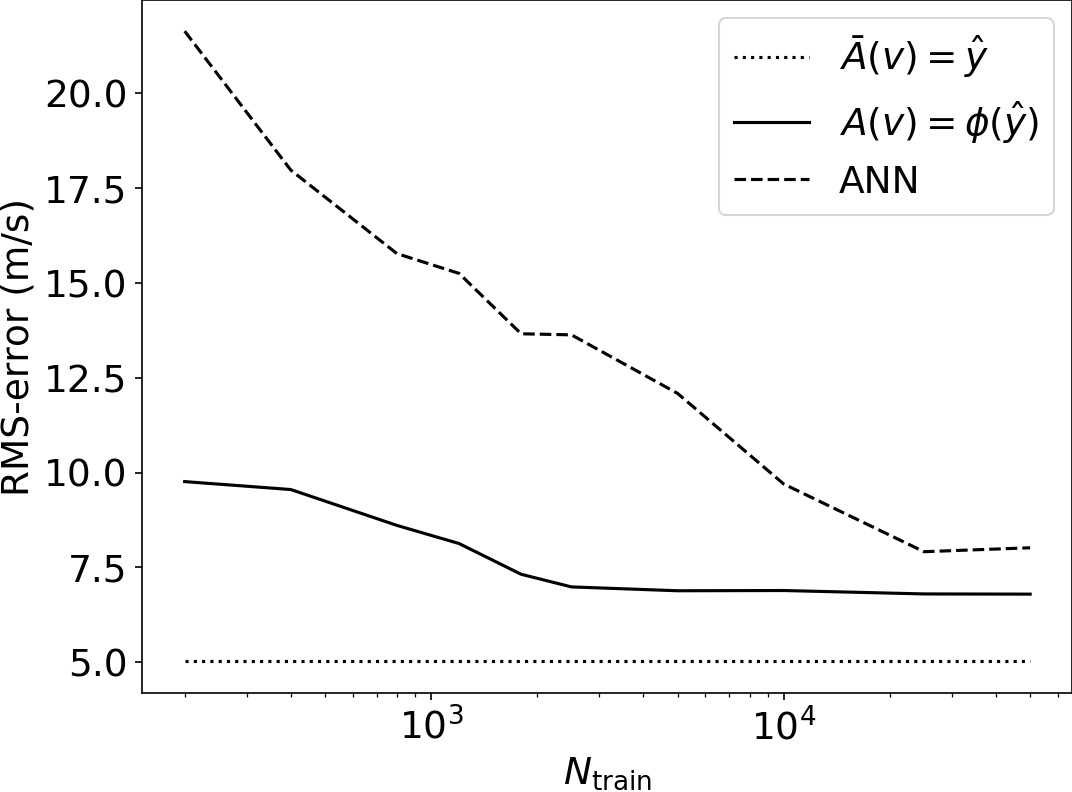}
  }
  \caption{Average RMS error of the NNAEE inversion, ANN based inversion, and the inversion using the accurate numerical model as a function of $N_{\mathrm{train}}$. The error of the NNAEE method is plotted with solid line, the error using the ANN inversion with dashed line, and the error using the accurate model with dotted line.}
  \label{fig:learning_curves}
\end{figure}

\subsection{Extrapolation capabilities}

{
In the first test cases, the phantoms were quite close to the phantoms used  in the training phase. While this is preferable, in reality  it is possible that the measured sample significantly differs from the training set samples. This type of case was simulated by creating a sample that differs from the training dataset. The sample contained star-shaped inclusions with the SOS between $1200\,\rm{m/s}$ and $1800\,\rm{m/s}$, and constant background with the SOS of $1500\,\rm{m/s}$, see Figure \ref{fig:reco5000_90000}. }

{
In the NNAEE and direct ANN reconstructions, the same trained ANNs as in the earlier test cases were utilized.
Examples of the reconstructed images and the corresponding errors are presented in Figure \ref{fig:reco5000_90000}. Detailed results can be found in Table \ref{tbl:result_90000} and Figure \ref{fig:learning_curves_90000}. The results show that the RMS-errors of the proposed NNAEE method are smaller than the  errors of the direct ANN inversion method and the CRM. On the other hand, the error levels of these methods are about two times larger compared to the first test case. With the NNAEE method the target is clearly visible, whereas with the other two approaches, the target can hardly be discerned. 
}

\begin{figure}[!ht]
{
  \center{
  \includegraphics[width=\textwidth]{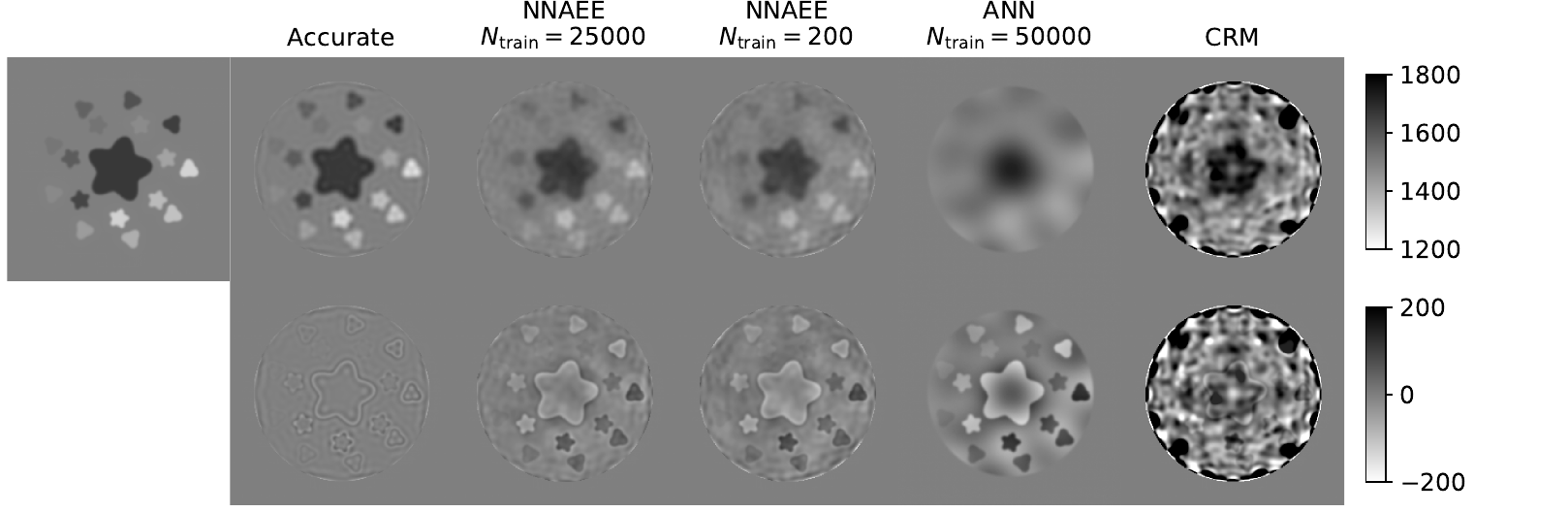}
  }
  \caption{Target is located on left, and reconstructions on upper row from left to right are computed with accurate model, NNAEE, with $N_{\mathrm{train}}=25000$, NNAEE, with $N_{\mathrm{train}}=200$, ANN based inversion, with $N_{\mathrm{train}}=50000$, and CRM. At lower row there are the corresponding error fields. The edge length of each rectangle is $0.215\rm{m}$, and the speed of sound and error are in $\rm{m}/\rm{s}$.}
  \label{fig:reco5000_90000}
  }
\end{figure}          

\begin{figure}[!ht]
{
  \center{
  \includegraphics[width=0.5\textwidth]{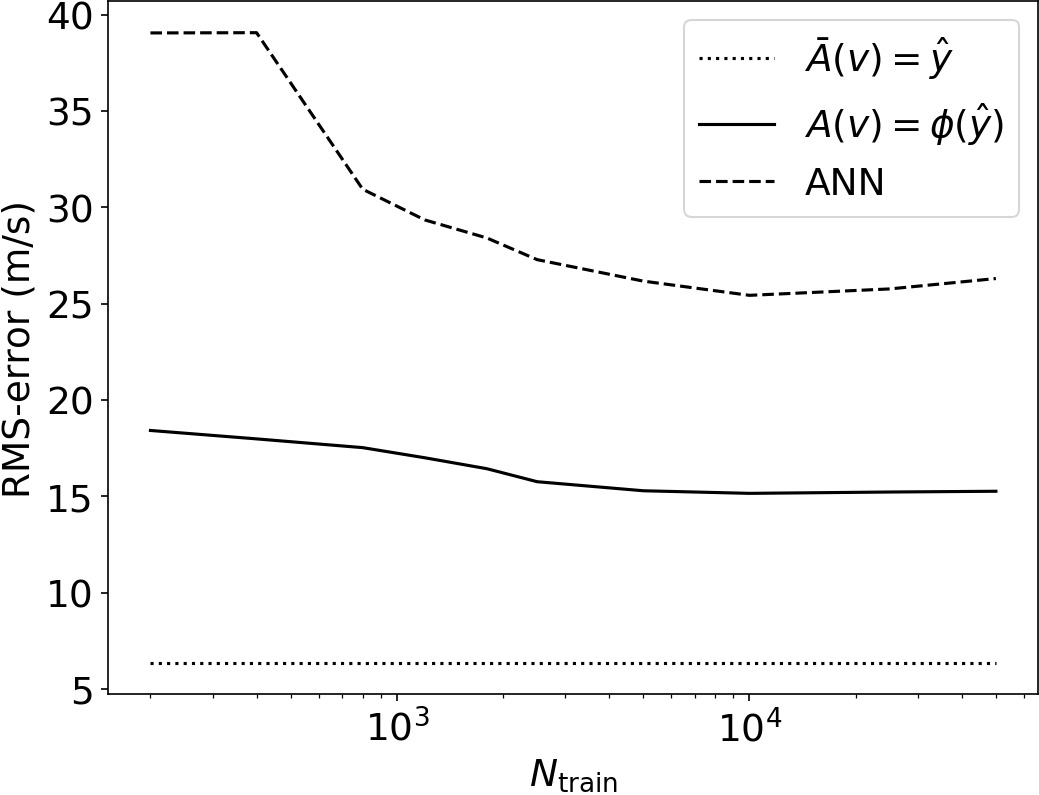}
  }
  \caption{RMS-error of the NNAEE inversion, ANN based inversion, and the inversion using the accurate numerical model, in $\rm{m}/\rm{s}$, as a function of $N_{\mathrm{train}}$. The error of the NNAEE method is plotted with solid line, the error using the ANN inversion with dashed line, and the error using the accurate model with dotted line.}
  \label{fig:learning_curves_90000}
}
\end{figure}

\begin{table}[!ht]
  \caption{RMS-errors of the inversions using NNAEE and ANN inversion as a function of $N_{\mathrm{train}}$ at the second test case. Error of the reconstruction using the accurate model, $\accurate{A}(\accurate{v})$, and, in addition, error of the reconstruction using the surrogate model $A(v)$, are presented as the lower limit and the upper limit reference value. Error levels of NNAEE are roughly half of the error levels compared to ANN based inversion, but much higher than the  error level of reconstruction using the accurate model in the reconstruction.} 
 \label{tbl:result_90000}
  \center{
    \begin{tabular}{|c|c|c|}
    \hline
    $N_{\mathrm{train}}$ & NNAEE error ($\rm{m}/\rm{s}$) & ANN error ($\rm{m}/\rm{s}$) \\ \hline \hline
    200 & 18.41 & 39.05\\ \hline
    400 & 17.98 & 39.07\\ \hline
    800 & 17.52 & 30.94\\ \hline
    1200 & 17.00 & 29.34\\ \hline
    1800 & 16.43 & 28.41\\ \hline
    2500 & 15.75 & 27.28\\ \hline
    5000 & 15.28 & 26.17\\ \hline
    10000 & 15.15 & 25.43\\ \hline
    25000 & 15.22 & 27.76\\ \hline
    50000 & 15.26 & 26.30\\ \hline \hline
    Accurate & 6.36 & \\ \hline
    Surrogate & 88.90& \\ \hline
  \end{tabular}
  }
\end{table}

\section{Conclusions}
\label{sec:concl}

In the current paper, we propose a method utilizing an ANN to compensate for the influence of modelling error, when using computationally light surrogate model in the reconstruction of SOS field in acoustic inverse problems. The method was {first} tested with 20 simulated measurements using different number of training datasets and the reconstructions were compared to an accurate reconstruction, CRM as well as to a conventional ANN inversion. The results show that the proposed NNAEE method can produce almost as good reconstructions as with the accurate model, while reducing the required reconstruction times significantly. In addition, its was shown that the proposed approach can work with fairly small number of training datasets. Increasing the size of the training set up to a certain limit, did not improve the quality of the reconstruction further. {In the second test, extrapolation capability of the method was tested with a phantom that was significantly different from the phantoms used in the training dataset. The results show that the proposed method outperformed CRM and ANN inversion also in this case.}

Further study is required for finding optimal hyperparameters of the ANN presenting $\varphi$. In addition, comparing the proposed method to other approximation error methods, like BAE{, and studying the effect of the training set quality on the method performance}, will be of interest. 

Even though, the method was demonstrated over acoustic inverse problem, we believe that it can be successfully utilized in other types of inverse problems.

\section*{Acknowledgments}

This research was supported by the Academy of Finland (the Finnish Center of Excellence of Inverse Modeling and Imaging, project 312344) and Academy of Finland project 321761.

\end{document}